\documentclass[superscriptaddress,aps,twocolumn]{revtex4-1}
\usepackage{bm}
\usepackage{float}
\usepackage{amssymb}
\usepackage{amsmath}
\usepackage{multirow}
\usepackage{lineno}
\usepackage{graphicx}
\usepackage{comment}
\usepackage{soul}
\usepackage[colorlinks=true,citecolor=blue,linkcolor=blue]{hyperref}
\usepackage[usenames]{color}
\usepackage{subfigure}

\begin{document}

\title{Intense high-harmonic optical vortices generated from a micro-plasma-waveguide irradiated by a circularly polarized laser pulse}

\author{Ke Hu}
\affiliation{Tsung-Dao Lee Institute, Shanghai Jiao Tong University, Shanghai 200240, China}
\author{Longqing Yi}
\thanks{lqyi@sjtu.edu.cn}
\affiliation{Tsung-Dao Lee Institute, Shanghai Jiao Tong University, Shanghai 200240, China}

\date{\today}

\begin{abstract}
A scheme for generating intense high-harmonic optical vortices is proposed. It relies on spin-orbit interaction of light when a relativistically-strong circularly polarized laser pulse irradiates a micro-plasma-waveguide.
The intense laser field drives a strong surface wave at the inner boundary of the waveguide, which leads to high-order harmonic generation as the laser traveling inside. For a circularly polarized drive laser, the optical chirality is imprinted to the surface wave, which facilitates conversion of spin angular momentum of the fundamental light into orbital angular momenta of the harmonics. A ``shaken waveguide" model is developed showing that the aforementioned phenomena arises due to nonlinear plasma response that modifies electromagnetic mode at high intensities.
We further show the phase velocities of all the harmonic beams are automatically matched to the driving laser, so that the harmonic intensities increase with propagation distance. The efficiency of harmonic production are related to the surface wave breaking effect, which can be significantly enhanced using a tightly focused laser. Our simulation suggests an overall conversion efficiency $\sim5\%$ can be achieved.
\end{abstract}
\maketitle

Light can carry spin and orbital angular momenta (SAM and OAM) \cite{Allen1992,Yao2011,Bliokh2015}.
The SAM is related to the handedness of circular polarization, and a photon can carry SAM
of $\sigma \hbar$, where $\sigma=-1$ or $+1$ for the left- or right-handed polarization, respectively.
The OAM is possessed by light beams that exhibit helical wave fronts,
which is typically expressed by a helical spatial phase $\exp(il\phi)$,
where $l$ is the topological charge and $\phi$ is the azimuthal angle.
Recently, high-order harmonic vortex sources have attracted considerable attention due to their
applications in many fields of science, such as optical
communication \cite{Gibson2004,Wang2012}, optical trapping \cite{Oneil2002},
and biophotonics \cite{Willig2006}.
Their special intensity profile makes them a useful
tool to control laser-matter interactions \cite{Onoda2004,Bliokh2008,Padgett2011,Vieira2018}.

Many studies have attempted to obtain intense, high-order vortex laser beams.
It has been experimentally proven that a
relativistic electron beam in a helical undulator can emit
high-order strong vortex radiation\cite{Hemsing2011, Hemsing2013}.
Besides, Compton backscattering of a twisted
laser from a gas target produces high-energy
photons with OAM \cite{Jentschura2011}.
Another type of  method that has promising perspective relies on
high harmonic generation (HHG) from high-power laser interacting with solid targets.
Such approaches typically rely on HHG via the relativistic oscillating mirror (ROM) mechanism \cite{Bulanov1994,Lichters1996,Baeva2006}, the driver can be a relativistic Laguerre-Gaussian beam \cite{Zhang2015,Denoeud2017},
a tightly focused circularly polarized (CP) pulse \cite{Wang20191, Zhang2021}, or a linearly polarized laser if plasma holograms are employed \cite{Leblanc2017}.
However, since the ROM mechanism is suppressed for CP drivers at normal incidence, it is challenging to generate intense circularly polarized vortex beams, that are of particular interest for controlling chiral structures \cite{Toyoda2012,Toyoda2013} and optical manipulation at relativistic intensities \cite{Wang20192}.

Recently, we demonstrated \cite{Yi2021} that high-harmonic optical vortices can also be generated when a high-power CP laser diffracts through a relativistic oscillating window (ROW). The laser drives chiral relativistic electron oscillations at the periphery of an aperture, which produces high-order harmonics via Doppler effect and facilitates spin-orbital interaction of light.
This could potentially boost the intensity of harmonic vortex beams since it no longer relies on the ROM mechanism. However, for typical plasma apertures (that can be produces by relativistic induced transparency \cite{Gonzalez-Izquierdo20161,Gonzalez-Izquierdo20162,Duff2020}, for example), which are a few times laser wavelength in diameter, only the edge of laser pulse near the periphery are responsible for HHG. This hinders achieving the goal of high laser-to-harmonics conversion efficiency.

In this letter, we show that the above obstacle can be tackled by irradiating a micro-plasma waveguide (MPW) with a relativistic CP laser.
The interactions of lasers with such targets have already shown
great potential in electron acceleration \cite{Snyder2019,Xiao2016}, radiation
generation \cite{Yi20161,Yi20162,Yi2019,Hu2021}, production of
ion beams \cite{Zou2015,Kluge2012} and manipulation of relativistic laser pulses \cite{Ji20161}.
Despite the progresses that have been made in both theory and experiment \cite{Snyder2019}, previous studies are mainly based on the linear theory of plasma waveguide \cite{Shen1991}, which
fails to capture the modification of electromagnetic mode due to the nonlinear plasma response on the boundary. Here, using first-order perturbation, we introduce a simple ``shaken waveguide" model, which suggests high-order harmonics are generated due to plasma oscillation on the inner wall, and the harmonic beams carry OAM if the drive laser is circularly polarized. Moreover, due to a novel phenomenon of self-phase-matching, this process continuously extracts energy from the drive laser, resulting in an overall conversion efficiency as high as $5\%$.

\begin{figure}[t]
\centering
\includegraphics[width=0.48\textwidth]{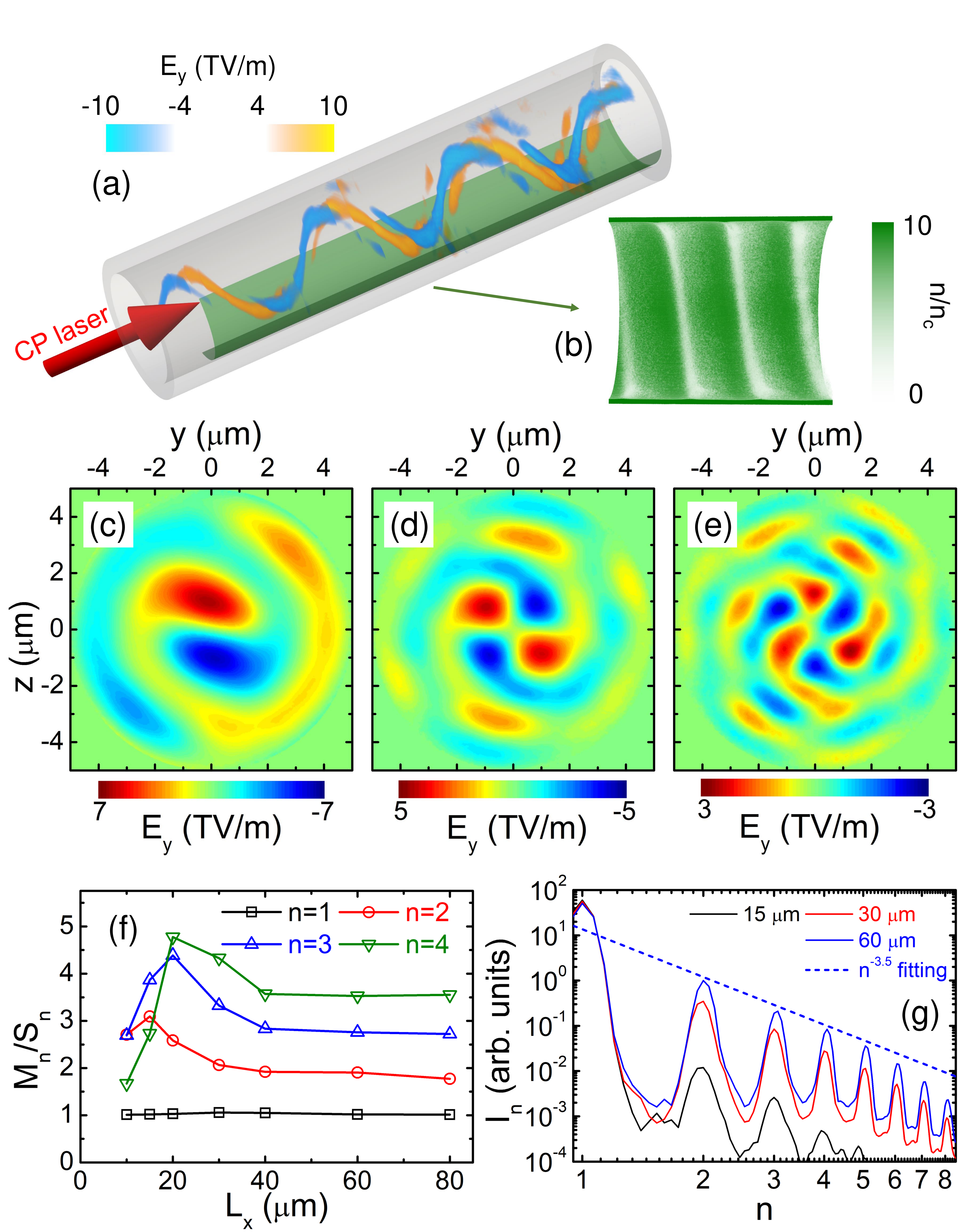}
\caption{
    (a) The 3D schematic setup of the proposed scheme. A circularly polarized laser pulse enters a MPW (grey tube) from the left, high-order harmonics are generated as the light bouncing between the channel walls, where (b) a co-propagating chiral surface wave driven by the laser is observed. The color-coded electric field $E_y$ in (a) shows a light spring structure of the harmonic beams ($\omega>2.5\omega_0$).
    (c-e) The $E_y$ field distributions on the $y-z$ cross-section, for the second-, third-, and fourth-order harmonics, observed at propagation distance $L_x=80\ {\rm \mu m}$, respectively.
    (f) The ratio of total angular momentum to spin angular momentum plotted as a function of propagation distance for harmonics of order $n=1-4$.
    (g) The spectra of the electromagnetic wave in the waveguide, observed at propagation distance of $15\ {\rm \mu m}$, $30\ {\rm \mu m}$ and $60\ {\rm \mu m}$.
    }
\end{figure}

We first demonstrate our scheme using a 3D particle-in-cell (PIC) simulation
with the EPOCH code \cite{Arber2015}. The setup of laser-waveguide interaction is illustrated
in Fig.~1(a).
A CP driving laser pulse enters a cylindrical waveguide along the $x$ axis from the left and travels to the right.
The simulation box has dimensions of
$x\times y\times z=16\times 14\times 14\ \mu{\rm m}^3$
and is sampled by $2400\times 560\times 560$ cells,
with four macroparticles for electrons and two for C$^{6+}$ per cell.
A moving window is used to improve computational efficiency, which follows
the propagation of the drive laser pulse in the MPW.
The laser field is $\mathbf{E}=(\mathbf{e_y}+i\sigma \mathbf{e_z})E_0 \text{exp}(-t^2/\tau_0^2)\text{exp}(ik_0x-i\omega_0t)$,
where $\mathbf{e_y}$ ($\mathbf{e_z}$) are the unit vectors in $y(z)$ direction, $E_0$ is the laser amplitude,
$\omega_0$ is the angular frequency, $k_0=2\pi/\lambda_0$ is the wavenumber, with
$\lambda_0=1\ \mu{\rm m}$ the laser wavelength, and $\tau_0=10.19\ {\rm fs}$, corresponding to
a temporal FWHM duration of $12.0\ {\rm fs}$.
The laser has a normalized amplitude $a_0=eE_0/mc\omega_0=16$, where $c$ is the speed of light, $m$ is the electron mass, and $e$ is the unit charge.
In this simulation, the laser is right-handed polarized ($\sigma=+1$).
The MPW target has a length of $L=100\ \mu{\rm m}$, an inner
radius of $r_0=5.0\ \mu{\rm m}$, and density of $n_0=30n_c$, where
$n_c=\epsilon_0m\omega_0^2/e^2$ is the critical density.

The laser energy are converted into waveguide modes when it enters the channel. Simultaneously, the relativistically-strong CP laser field drives collective electron oscillations, forming a chiral surface wave that co-propagates with the laser, as shown in Fig.~1(b).
As the light bounces between such oscillating surfaces, it experiences a relativistic Doppler shift, which can lead to the generation of high-order harmonics \cite{Bulanov1994,Baeva2006}.

The three-dimensional harmonic $E_y$ fields ($\omega>2.5\omega_0$) are shown in Fig.~1(a), which presents a helical structure known as light spring \cite{Pariente2015}. This indicates that each harmonic frequency is associated to a spatial LG-like mode, and the topological number $l(\omega)$ varies linearly with $\omega$. Such a structure is of particular interest for controlling the topology of laser-plasma accelerators \cite{Vieira2018}.

In Figs.~1(c-e), we show the field distribution of second, third, and fourth harmonics in the cross-section perpendicular to the propagation axis. One can see a relation $l=(n-1)\sigma$ is satisfied in all cases, which is further confirmed by Fig.~1(f), where the ratios of longitudinal component of the total angular momentum $M_n = \varepsilon_0[\int\mathbf{r} \times (\mathbf{E}_n \times \mathbf{B}_n)d\mathbf{r}]_x$ and the SAM $S_n = W_n/n\omega_0$ of the $n$th harmonic are plotted against propagation distance ($L_x$) for $n = 1-4$, where $\mathbf{E}_n$, $\mathbf{B}_n$ and $W_n$ denote the electric field, magnetic field and energy of the $n$th harmonic, respectively. Physically, this relation ensures the conservation of energy and angular momentum during the HHG process: the sum of SAM of $n$ fundamental-harmonic photons ($n\times\sigma\hbar$) are transformed into the SAM ($\sigma\hbar$) and OAM [$(n-1)\sigma\hbar$] of one $n$th-harmonic photon. We note that the left- or right-handedness of the drive CP laser only changes the sign of $l$ for the generated harmonic vortices and has little effects on other results of this work, we therefore consider only the right-handed case $\sigma = +1$ hereafter.

The harmonic spectra of the electromagnetic field observed at different propagation distance inside the MPW is illustrated in Fig.~1(g), the harmonic beams continuously extract energy from the driver for a few tens of microns. The HHG efficiency, defined as the energy of the harmonic beams ($n\geq2$) over total energy of drive laser, is $3.64\%$. The spectra can be fitted with a power-law shape $I_n\propto n^{-3.5}$, where $I_n$ is the intensity of the $n$th harmonics. This indicates the HHG mechanism is very similar to that of ROW \cite{Yi2021}. This is expected because a waveguide can be considered as a train of co-axial diffraction apertures with the same radii. However, in our scheme, the chiral surface electron oscillation forms a co-propagating surface wave, which is crucial for achieving a high conversion efficiency, as it allows for a long-time interaction between the drive laser and the oscillating surface electron layer.

The fact that harmonic intensities grow with propagation distance implies they have the same phase velocity as the driver, such that harmonics generated at different longitudinal positions can add constructively. This is non-trivial according to the linear theory of plasma waveguide \cite{Shen1991,Yi20161}.
The phase velocity is $v_{ph,n} = n\omega_0/k_{x,n}$, where $k_{x,n} = \sqrt{n^2k_0^2-k_{T,n}^2}$, with $k_{T,n} = x_{1,n}/r_0$  the transverse wave number and $x_{1,n}$ the first root of eigenvalue equation (here we consider only fundamental waveguide modes). The subscript $n$ denotes $n$th harmonic. Since $x_{1,n}$ does not increase proportionally to $n$, in fact it changes little with varying frequency \cite{Shen1991}, one would expect different orders of the harmonic beams would propagate at different velocities. As a result, the difference in phase velocities between the fundamental light and a high-order ($n>10$) harmonic would potentially cause catastrophic negative interference within few micrometers, which conflicts with the numerical results.\\

To explain the observed phenomena, we introduce a ``shaken waveguide" model, taking into account the co-propagating surface wave as a periodical deformation of the MPW. We take the linear plasma waveguide theory as the zeroth-order approximation. The radial component of electric field, which is responsible for deforming the MPW, is

\begin{equation}
E_r(x,r,\phi)=E_0J_0(k_Tr){\rm exp}(i\sigma\phi){\rm exp}(ik_xt-i\omega_0t),
\end{equation}
where $r=\sqrt{y^2+z^2}$ is the radial coordinate, $J_0$ is the
Bessel function of the first kind, $k_x$ and $k_T$ are longitudinal and
transverse wavenumber components of the drive laser, respectively.
The subscript $n=1$ is omitted for simplicity.

As the first-order approximation, we assume the cross section of the MPW remains circular, and its center is shifted from the original MPW axis by $\delta r=-\delta r_0{\rm exp}(i\sigma\phi){\rm exp}(ik_xt-i\omega_0t)$ due to collective electron oscillation, where the phase of the electron motion is the opposite to that in Eq.~(1). The amplitude of this displacement $\delta r_0$ is assumed to be small, such that a perturbative method can be applied. Submitting this running periodic displacement of the waveguide centre back into Eq.~(1) with $r\rightarrow r-\delta r$ (similarly to the expression for the azimuthal electric component $E_\phi$), and using a Taylor expansion and the Jacobi-Anger identity \cite{Cuyt2008}, the transverse electric field can be written as a sum of harmonics,
\begin{equation}
\begin{aligned}
\mathbf{E}_{\perp}(x,r,\phi)\propto&(\mathbf{e_y}+i\sigma \mathbf{e_z})\sum_{n=1}^{\infty} E_0\times\\
&{\rm exp}[ink_xx-in\omega_0t+i(n-1)\sigma\phi].
\end{aligned}
\end{equation}
We note that the intensities of harmonic beams can not be obtained by the first-order perturbation unless the nonlinear term is small, i.e. $k_T\delta r_0\ll1$, this is not always true for the laser-plasma parameters under consideration. However, here we are mainly interested in the phase term of electromagnetic waves in the MPW. It could interpret the spin-orbital interaction of light occurs during HHG, as well as the phase velocity modification of the harmonic beams. Because the underlying physics, namely the conservation law of angular momentum and the change of boundary condition, must always apply regardless the strength of the nonlinearity.

One can see from Eq.~(2) that the generated harmonic beams are circularly polarized with the same handedness as the driver, and have helical phase fronts ${\rm exp}[i\sigma(n-1)\phi]$, from which the relation $l=(n-1)\sigma$ is recoverd.

\begin{figure}[t]
\centering
\includegraphics[width=0.40\textwidth]{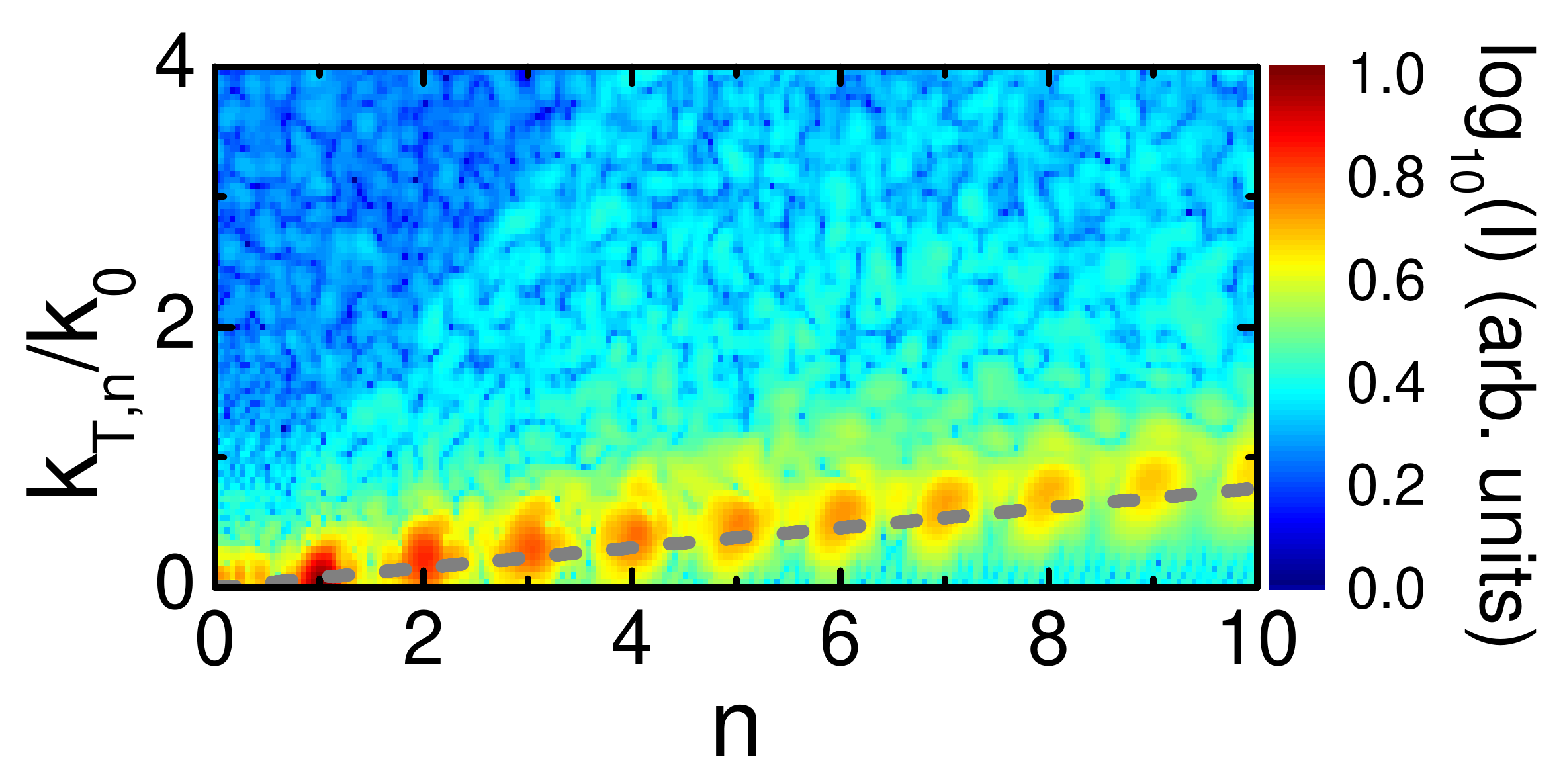}
\caption{Two-dimensional Fourier diagram of the $E_y$ field in the waveguide, showing the intensity distribution of the generated harmonic beams in the $k_T-\omega$ space.
  The grey dashed line is obtained by $k_{T,n}/k_0=nx_1\lambda_0/2\pi r_c$.}
\end{figure}

The deformation of MPW also modifies the boundary conditions, and consequently the propagation of the harmonic beams. In a ``shaken waveguide", a longitudinal density modulation exists on the inner surface, with a period equals to the surface wave $2\pi k_x^{-1}$.
Notably, Eq.~(2) suggests the longitudinal wavenumber for $n$th harmonic beam is $k_{x,n} = nk_x$, thus integer number ($n$) of harmonic cycles exist between one surface wave crest [dark parts in Fig. 1(b)]. In this way, the dynamical boundary condition is satisfied in all space and time, and therefore the phase velocities of all harmonic beams are locked with the driver.

This novel phenomenon of self-phase-matching is one of the key findings of this work, which allows for a highly-efficient production of the high-harmonic optical vortices. The result is further examined with high-resolution 2D PIC simulations ($dx\times dy=\lambda_0/400\times \lambda_0/200$). All the laser-plasma parameters are kept same as the 3D simulation presented in Fig.~1. We perform 2D Fourier analysis for the $E_y$ field observed at $L_x = 80\ {\rm \mu m}$ in the $y-t$ space. The resulted intensity distribution in $k_T-\omega$ phase space shows a linear relation $k_{T,n} = nk_{T,1}$, which proves that all harmonics are propagating with the same phase velocity as the driving laser.\\

\begin{figure}[t]
\centering
\includegraphics[width=0.48\textwidth]{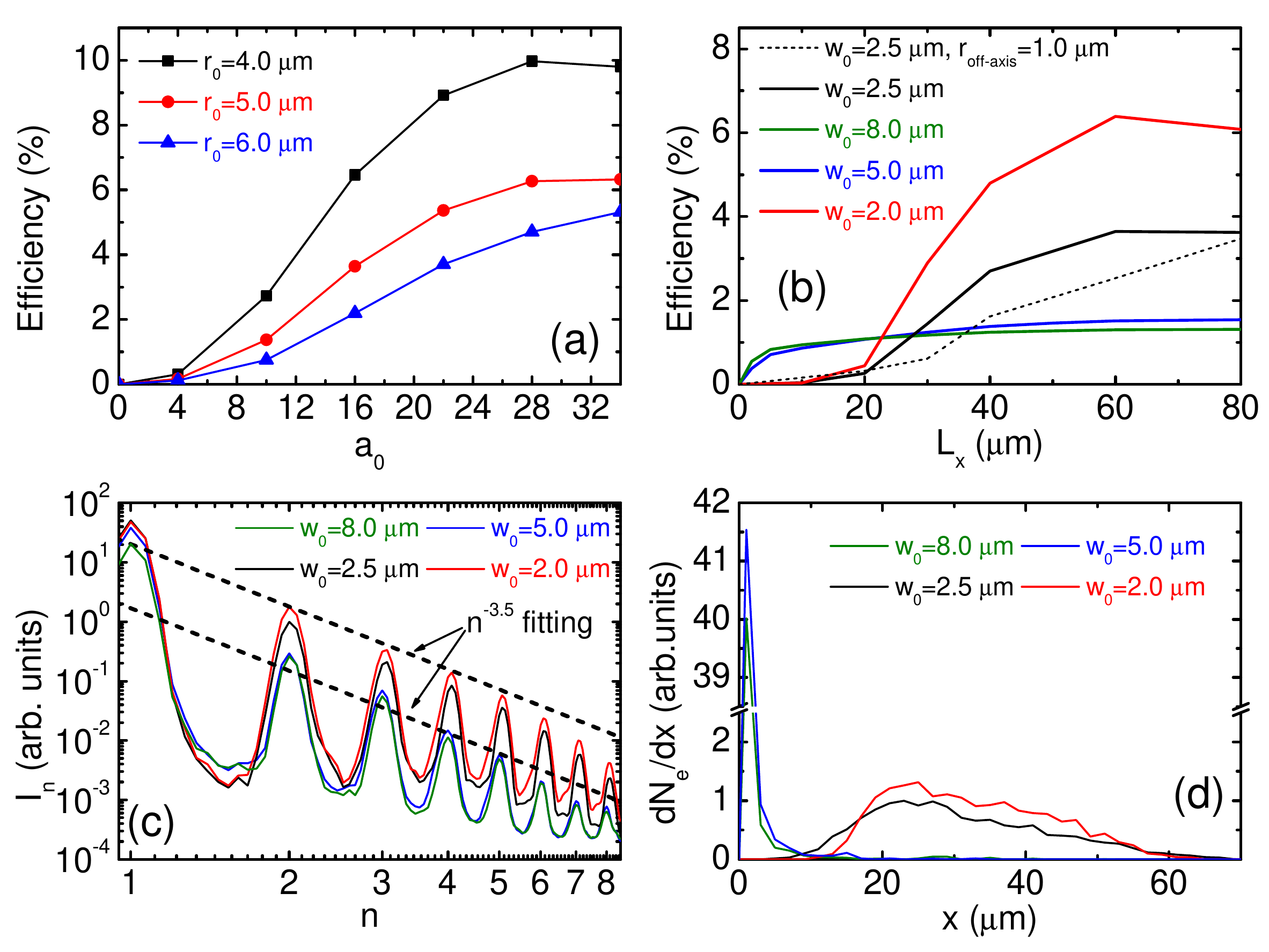}
\caption{
(a) The HHG efficiency plotted as a function of $a_0$ for different $r_0$, where the laser waist $w_0$ is fixed at $r_0/2$.
(b)  The HHG efficiency are plotted against propagation distance for a laser beam focused to different waist $w_0$ at the entrance of a MPW.
The black dashed line shows the case of drive laser mis-alignment by 1 ${\rm \mu m}$ off axis.
The harmonic spectra observed at $L_x = 60\ {\rm \mu m}$ are presented in (c), and the histogram of the initial positions for the escaped high-energy electrons is shown in (d), the longitudinal position with high electron number count indicates where the SWB effect takes places. In (b-d)  the laser energy is fixed to be $0.88$ {\rm mJ} and the channel radius is $r_0=5.0\ {\rm \mu m}$ for all cases.
}
\end{figure}

In the following, we investigate the parameter dependence of the proposed HHG scheme, and seek to improve the overall conversion efficiency for a given laser energy and a fixed MPW. The study is performed with 3D PIC simulations, the parameters are the same as in Fig.~1 unless otherwise stated.

The general results are summarized in Fig.~3(a), where the ratio of laser waist and channel radius is set to be $w_0/r_0 = 0.5$ (discuss below). One can see that when the laser is weak, the HHG efficiency increases rapidly with $a_0$, then the slope becomes almost linear until it saturates at $5\%-10\%$.
This trend is true for all the cases with different MPW, however the simulations suggest for the waveguides with smaller radii,
the efficiency slopes are steeper, saturate at lower driving intensities, and typically slightly higher HHG efficiencies are obtained.

For practical reasons, it is of particular interest to optimize the drive laser for a given energy so that the harmonic vortices can be efficiently produced. Interestingly, we found the HHG efficiency changes dramatically with varying waist-to-radius ratio $w_0/r_0$ in this situation.

The results are presented in Fig.~3(b), where four simulations are performed with laser waist $w_0=2.0\ \rm \mu m$, $2.5\ \rm \mu m$, $5.0\ \rm \mu m$, and $8.0\ \rm \mu m$, the drive energy is the same (the laser $a_0$ are adjusted accordingly) and the MPW radius is fixed at $r_0 = 5\ \rm \mu m$. One can see that when $w_0\geq r_0$, almost all harmonic are generated near the entrance, the HHG intensities increase rapidly within the first a few microns as the laser enters MPW, but the HHG process terminates very soon ($L_x<10\ \rm \mu m$), the energy of harmonic beams remains the same afterwards, and the efficiency is $\sim1\%$. On the other hand, when $w_0<r_0$, the harmonics are mainly generated inside the channel, the intensities rise slowly in the beginning, but the process lasts for a much longer distance ($L_x\sim60\mu m$), the HHG efficiency reaches $\sim5\%$ in the end. We therefore conclude that a tightly focused laser can significantly enhance the HHG efficiency for the proposed scheme.
Such a setup can be challenging for the-state-of-the-art laser systems due to misalignment issue. Fortunately, our numerical results suggest the final HHG efficiency drops little due to a misalignment on the order of magnitude of $1 {\rm \mu m}$, as shown by the black dashed line in Fig.~3(b).
Such requirement of  pointing stability has been achieved in recent laser-waveguide experiments \cite{Snyder2019}.

The HHG spectra for the cases presented in Fig.~3(b) are shown in Fig.~3(c). Apparently, they all have a power-law shape that can be fitted with $I_n\propto n^{-3.5}$, while the harmonics produced by tightly focused drive lasers are several times stronger.

Since the high-order harmonic vortices are generated through a similar mechanism with the ROW \cite{Yi2021}, namely the chiral electron oscillation on the periphery of a high-power laser beam,
the HHG efficiency depends crucially on the amplitude of the oscillation.
A large amplitude is associated with the surface wave breaking (SWB) effect, i.e. when the drive laser is sufficiently strong, some of the oscillating electrons acquire enough energy to escape, they are ejected into the vacuum as energetic micro-bunches \cite{Naumova2004}. When SWB occurs, the surface wave is highly nonlinear, and its amplitude is at maximum, thus highly-efficient HHG is expected. However, unlike the ROW, in our scheme the harmonic vortices are generated gradually as the laser propagating in the MPW, therefore achieving a high HHG efficiency requires the SWB effect to take place over a significant portion of the propagation distance.

How long the SWB effect can last is determined by the number of the electrons escaped.
As more and more high-energy electrons are ejected into the vacuum, it in turn builds up a Coulomb barrier that prevent further electrons from escaping \cite{Yi2019,Hu2021}. This ultimately suppresses SWB, leading to a reduction of the surface wave amplitude, and consequently the HHG efficiency. In order to examine this effect, for each cases presented in Fig.~3(b), we track the trajectories of high-energy electrons ($\gamma>10$) obtained at the end of the MPW, and plot the histogram of their injection positions in Fig.~3(d). One can see that in all cases, the time that high-harmonic intensities rising rapidly coincides with the phase with SWB-induced electron injection, after which both the electron emission and the HHG process stops.

In the cases of $w_0=5.0\ {\rm \mu m}$ and $8.0\ {\rm \mu m}$, almost all energetic electrons are injected at the vicinity of the waveguide entrance. The strong impact of laser pulse with the MPW front surface produces a large number of electrons. This effect is dominant for electron injection as long as the normalized laser intensity at the rim of MPW is $a(r=r_0)\geq1$. In this case, the Coulomb barrier is established within several micrometers, leading to a premature quenching of the HHG process. It should be noted that the number of the high-energy electrons contribute little to the harmonic production, the electron emission is merely a signature of the SWB effect, it is the distance that electrons continue to emit that plays a vital role.

On the contrast, for $w_0=2.0\ {\rm \mu m}$ and $2.5\ {\rm \mu m}$, both cases satisfy that $a(r=r_0)\ll 1$, the number of high-energy electrons originated from the entrance is negligibly small. The electron injection starts at around $L_x=10.0\ {\rm \mu m}$, and lasts up to $L_x=60\ {\rm \mu m}$, during which the high-harmonic vortices are generated efficiently due to the lasting SWB effect.\\

In conclusion, we demonstrate that high-harmonic vortices can be generated with high overall efficiency up to $\sim5\%$ via ultra-intense CP laser irradiating a micro-plasma-waveguide.
A ``shaken waveguide" model is introduced, which suggests spin-orbital interaction of light takes place in the high-harmonic generation process when the drive laser bounces between the channel walls, where a co-propagating chiral surface wave is presented.
The surface wave also modifies the boundary condition so that all the generated harmonic beam propagates with the same velocity as the drive laser in the waveguide. This allows for constructive interference of the harmonic lights generated along the propagation path.
We show that a tightly focused laser beam is preferred for achieving a high overall efficiency, as it avoids massive electron injection near the entrance of waveguide, therefore the amplitude of surface wave can remain high for a long distance.
Our study paves the way to a powerful laser-plasma-based high-harmonic vortex source, and provides theoretical insights into laser-waveguide interaction at ultra-high intensities.

\begin{acknowledgments}
This work was supported by the National Natural Science Fund for Excellent Young Scientists Fund Program (Overseas), the Shanghai Pujiang Talent Plan (No. 21Z510104677), and the National Key R$\&$D Program of China (No. 2021YFA1601700).
\end{acknowledgments}



\end{document}